\documentclass[a4paper]{aa} 
\usepackage{isolatin1}      
\usepackage{graphicx}
\usepackage{natbib}
\begin{document}
   \title{The runaway black hole GRO J1655-40\thanks{Based
on observations with the NASA/ESA Hubble Space Telescope,
obtained at the Space Telescope Science Institute, which is
operated by AURA, Inc.  under contract No NAS 5-26555}
}
   \author{I.F. Mirabel \inst{1,2}
	\and
	R. Mignani \inst{3}
	\and
	I. Rodrigues \inst{1}
	\and
        \\
	J.A. Combi \inst{4}
	\and
	L.F. Rodr\'\i guez \inst{5}
	\and
        F. Guglielmetti \inst{6,7}
	}

   \offprints{I. F. Mirabel, \email{fmirabel@cea.fr} }

   \institute{
	Service d'Astrophysique / CEA-Saclay, 91191 Gif-sur-Yvette, France
	\and
	Instituto de Astronomía y Física del Espacio/Conicet, Argentina
	\and
	European Southern Observatory, Karl-Schwarzschils-Strasse 2,
		Garching bei M\"unchen DE-85740, Germany 
	\and
	Instituto Argentino de Radioastronomía, C.C.5, (1894) Villa
		Elisa, Buenos Aires, Argentina
	\and
	Instituto de Astronomía, UNAM, Apartado Postal 3-72, 58089
	Morelia, Michoacán, México 
	\and
       Max Planck Institute f\"ur Extraterrestrische Physik, Giessenbachstrasse, Postfach 1312, D-85748, Garching, Germany  
	 \and
          Max Planck Institute f\"ur Plasmaphysik, 
Boltzmannstrasse 2, D-85748, Garching, Germany        
        }

   \date{{\bf {\large Published in \aap, Vol. 395, pages 595--599}}}

   \authorrunning{I.F. Mirabel et al.}

   \titlerunning{The runaway black hole GRO J1655-40}

   \abstract{   We have used the Hubble Space Telescope to measure the 
motion in the sky and compute the galactocentric orbit of the black hole 
X-ray binary GRO J1655-40. The system moves with a runaway space velocity 
of $112\pm 18$ km s$^{-1}$ in a highly eccentric ($e = 0.34\pm 0.05$)
orbit. The black hole was formed in the disk at a distance greater 
than 3 kpc from the Galactic centre and must have been shot to 
such eccentric orbit by the explosion of the progenitor star. 
The runaway linear momentum and kinetic energy of this black hole binary  
are comparable to those of solitary neutron stars and
millisecond pulsars. GRO J1655-40 is the first black hole for which 
there is evidence for a runaway motion imparted by a natal kick 
in a supernova explosion.
   }

   \maketitle
%

\section{Introduction}

Neutron stars are known to have large transverse motions on the plane
of the sky which are believed to result from natal kicks imparted by
supernova explosions.  Energetic explosions have also been invoked in
models of the core collapse of massive stars onto black
holes. However, there have been few observations that constrain models
of the physical processes by which stellar-mass black holes are
formed.  The measurement of a large radial velocity for the centre of
mass of the black hole X-ray binary GRO J1655-40 \citep{Orosz,
Shahbaz}, together with chemical elements found by \cite{Israelian} in
the surface of the donor star, provided observational support to the
idea that black holes -as neutron stars- may form in supernova
explosions that impart strong natal kicks to the collapsed objects.

If a black hole is accompanied by a mass-donor star, it is possible to
determine the radial velocity, proper motion, and distance of the
system, from which one can derive the space velocity, track the path
to the site of birth, and constrain the strength of the natal
kick. Presently, the most accurate proper motions of X-ray binaries
are obtained following at radio wavelengths with Very Long Baseline
Interferometry (VLBI) the motion in the sky of the associated compact
microquasar jets, as done recently for the halo black hole binary XTE
J1118+480 \citep{Mirabelnat}. This has not been possible for GRO
J1655-40 because there is no VLBI calibrator nearby, and in the last
years the radio counterpart faded away below detection limit. In this
context, we carried out relative astrometry of the secondary star
using optical images obtained with the Hubble Space Telescope 6.3
years apart. Here we report the proper motion of GRO J1655-40 which together 
with the radial velocity allows us -taking into account the uncertainties 
in the distance- to determine the parameters of its runaway kinematics 
and galactocentric orbit.

\section{The proper motion of GRO J1655-40}

\begin{figure*}[htb]
{\centering
\resizebox*{0.8\textwidth}{!}{\rotatebox{0}{\includegraphics{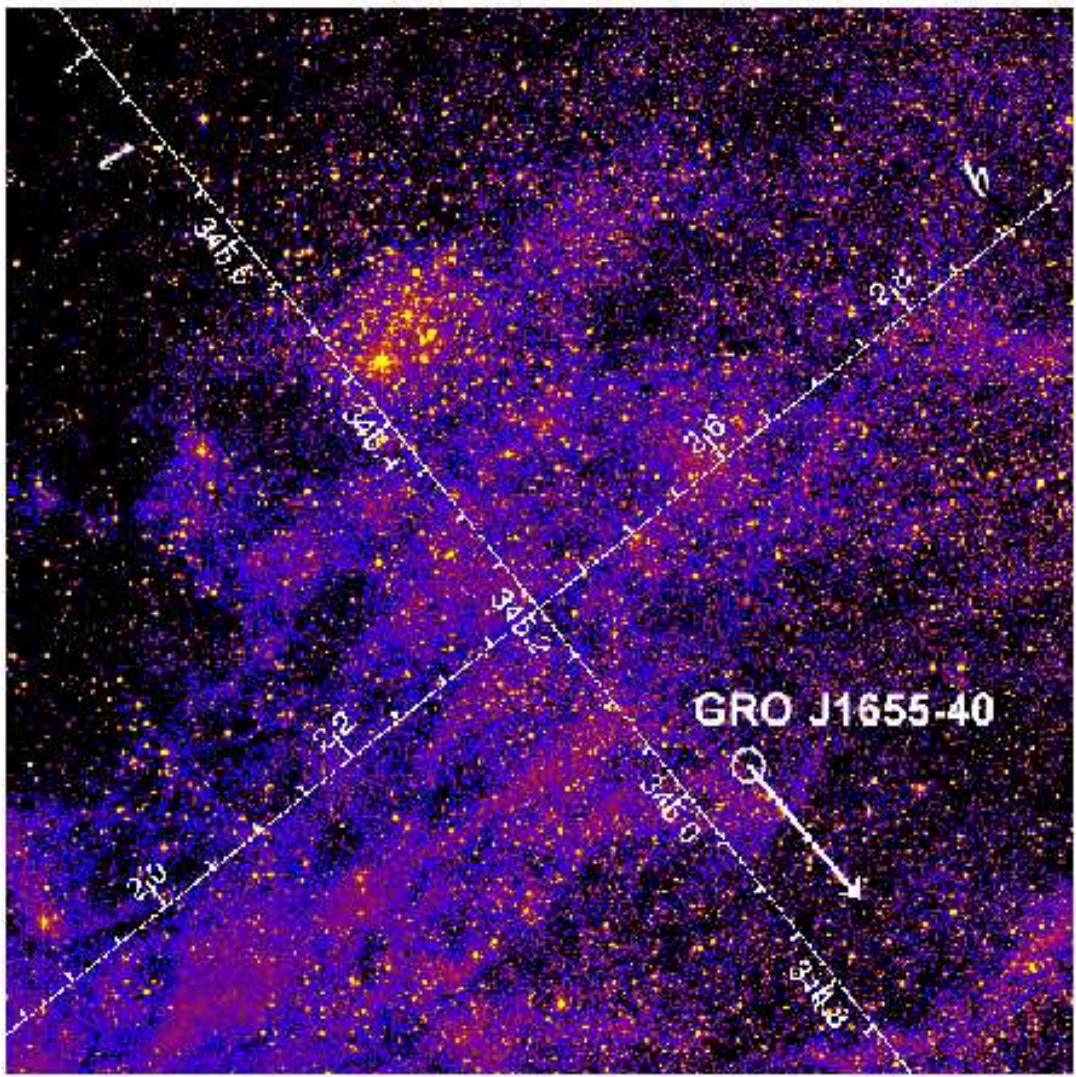}}}
\par}
\caption{\textbf{Position of GRO J1655-40 
on a R band image from the Digitized Palomar Observatory Sky Survey II 
(POSS II), in Galactic coordinates.} The arrow shows the direction of 
the motion at
a rate of 5.2 $\pm$ 0.5 mas yr$^{-1}$ measured with the Hubble Space 
Telescope.  Most of the stars around 
{\it l} = 345.44$^{\circ}$, {\it b} = + 2.43$^{\circ}$ belong to 
the open cluster NGC 6242 which is at a distance of 1 $\pm$ 0.1 kpc 
from the sun. Because of the uncertainty in the distance to GRO J1655-40  
the association with the cluster cannot be assessed.
\label{GROpath}}
\end{figure*}

On April 1996 the field of GRO J1655-40 was observed for the first
time with the WFPC2 on the Hubble Space Telescope \citep{Tavani}.  In
quiescence, the secondary star has an apparent magnitude of m$_{V}$
$\sim$ 17.2. The target had been placed at the center of the
Planetary Camera (PC) chip, with a pixel size of $0\farcs045$.
Observations were obtained through the 675W filter ($\lambda =
6717$ \AA; $\Delta \lambda = 736$ \AA). A total of 16 exposures of 40
s each were acquired on April 26$^{th}$ and the same sequence was
repeated on April 27$^{th}$ but with a different telescope roll angle.
Both sets of exposures were taken in groups of 4 and each group was
dithered by a few pixels in both Right Ascension and Declination with
respect to the others.

On June 20$^{th}$ 2001 a total of 18 exposures of 40 s each were
acquired by us, with the same observational set-up as in April 1996,
but without any dithering between single exposures.  After a
pre-reduction with the standard HST pipeline reduction (debiasing,
dark removal, flatfielding), groups of well aligned exposures (i.e.
with relative shift smaller than 0.01 pixel) were combined with a
median filter, stacked and cosmic ray hits filtered out.

We finally ended up with 4 images for each of the two April 1996
observations and one image for the June 2001 one.  All the 1996 images
were then registered on the 2001 one, previously aligned in Right
Ascension and Declination, by fitting the coordinate transformation
between grids of reference objects.  The correction for the WFPC2
geometrical distortions, optimized for the used filter \citep{Trauger}
was taken into account.  The whole procedure was iterated until the
transformation residuals were all below $\simeq 1.5~\sigma $.  We
finally computed the relative displacements of our target for each of
the 8, independent, couples of 1996/2001 images and we averaged the
results. The computed proper motion is $\mu_{\alpha} = -3.3 \pm 0.5$
mas yr$^{-1}$ and $\mu_{\delta} = -4.0 \pm 0.4$ mas yr$^{-1}$,
corresponding to an overall yearly displacement $\mu = 5.2
\pm 0.5$ mas yr$^{-1}$ along a Position Angle of $220^{\circ} \pm
5^{\circ}$. At a distance D(kpc) this proper motion corresponds to a
transverse velocity on the plane of the sky of (25$\pm$3) D(kpc) km
s$^{-1}$. In Figure~\ref{GROpath} we show the path of the black hole
binary on a $R$-band image of the Digitized Palomar Observatory Sky
Survey II (POSS II). NGC 6242 is a well studied open cluster at a
distance of $1020\pm 100$ pc from the Sun
\citep{Glushkova}. Figure~\ref{GROpath} shows that GRO J1655-40 is
close to the boundary of a dark cloud \citep[cataloged as DC344.9+2.6
in][]{Hartley}, which could explain the relative large reddening of
the secondary star in GRO J1655-40.

\section{The distance of GRO J1655-40}

In order to use our proper motion measurement to constrain the space
velocity of the black hole, we need to know its
distance. Unfortunately, as for most X-ray binaries, the distance to
GRO J1655-40 is rather uncertain. In the following we discuss the
observational constrains to the distance of this source.

\subsection{The relativistic distance}

It is widely believed that the distance of GRO J1655-40 was well
determined solely from the kinematics of the two sided radio
jets. This is not true.  The relativistic time delay of the motion of
the ejecta in the sky is given by the two equations
\citep{Mirabelnat94}:

\begin{equation}
\mu_{r,a} = \frac{\beta sin(\theta)}{1 \pm \beta cos(\theta)}
\frac{c}{D}  , 
\label{eq_rel}
\end{equation}

\noindent where $\mu_{r}$ and $\mu_{a}$ are the proper motions of the
receding and approaching jets. In this system of two equations there
are three unknowns: the angle with the line of sight of the jet axis
$\theta$, the velocity of the jets $\beta = \frac{v}{c}$, and the
distance D. We point out that using solely the observations at radio
wavelengths by \cite{Hjellming} it is not possible to solve these
equations, unless one assumes a value for one of the three variables
$\theta$, $\beta$, or D. \cite{Hjellming} assumed $\theta =
85^{\circ}\pm2^{\circ}$, from which one derives $\beta = \frac{v}{c} =
0.92$ and D = 3.2 kpc. As already noticed by \cite{Orosz}, in this
case the axis of the jet and the axis of the orbital plane differ by
$\sim 15^{\circ} \pm 2^{\circ}$.

We point out that the assumption that the jet axis is parallel to the
axis of the orbital plane is equally consistent with the observations
at radio wavelengths. The jet axis and the orbital plane must be
coupled, since the period of rotation of the jets about the jet axis
\citep{Hjellming} is -within the uncertainties-, the same as the 2.6
day orbital period of the binary \citep{Orosz}. If one assumes that
the twin jets are perpendicular to the orbital plane of the binary,
from $\mu_{r} = $ 45 mas/day and $\mu_{a} = 54$ mas/day
\citep{Hjellming}, $\theta = 70.2^{\circ}\pm1.9^{\circ}$
\citep{Greene} result $\beta = \frac{v}{c} = 0.27 \pm 0.03$ (where $v$
is the velocity of the jets and c the speed of light) and a distance
$\mathrm{D} = 893 \pm 100$ pc. Under this assumption, the distance
would be a factor 3.5 closer than commonly assumed, the jet velocity
would be similar to that in SS433, and GRO J1655-40 would not be a
superluminal source.

In summary, from the data at radio wavelengths and the system of two
equations (\ref{eq_rel}) with three unknowns one can only derive with
certainty a relativistic upper limit \citep{Mirabelrev} given by
$\mathrm{D} \leq \mathrm{c} / (\mu_{r} \mu_{a} )^{-1/2} = 3.5$ kpc.

\subsection{The distance and the interstellar matter along the line of sight}

A distance for GRO J1655-40 was proposed on the basis of optical
\citep{Bailyn2} and X-ray \citep{Greiner,Ueda,UedaErr} measurements of
the column of interstellar matter in the line of sight, under the
assumption that the absorbing material is distributed homogeneously
between the source and the observer.  However, GRO J1655-40 is at
relatively high Galactic latitude (Galactic longitude and latitude $l
= 345.0^{\circ}$, $b = +2.2^{\circ}$) in the Scorpius region of the
sky which contains rather clumpy optical dark clouds in the foreground
(see Figure \ref{GROpath}), that have $60 \mu$m and $100 \mu$m IRAS
counterparts of dust emission. From a study of the reddening undergone
by the stars that are at distances between 700 pc and 1900 pc, it is
known that most of the reddening in this region of the sky occurs in
the local arm within 700 pc from the Sun \citep{Crawford}.

On the other hand, a kinematic distance was proposed from the radial
velocity of absorption features in the HI $\lambda$21cm line spectrum
\citep{Tingay}. However, it is known that in this region of the sky 
at distances $\leq 1900$ pc there are clouds with anomalous velocities
of up to $-50$ km s$^{-1}$ \citep{Crawford}. Therefore, it is not
possible to derive the distance of GRO J1655-40 only from the column
density and/or kinematics of the interstellar matter in the line of
sight.

\subsection{Constrains on the distance from the properties of the
secondary star}

The main argument in favor of the canonical distance of $\sim$3.2 kpc
is based on the flux, color and size of the secondary. The radius is
inferred from the photometric light curves which provide evidence that
the secondary fills its Roche lobe \citep{Orosz, Shahbaz, Soria,
vanderHooft, vanderHooft2, Phillips}. From the optical spectrum
\citep{Orosz} and interstellar absorption
\citep{Horne} a temperature is derived, which together 
with the radius provides an intrinsic luminosity. In the most recent
model by \cite{Beer} the secondary would have a luminosity of 21 $\pm$
6.0 L$_{\odot}$, which is consistent with a distance $\geq$2 kpc.
These authors estimate masses for the black hole M$_{BH} = 
5.4 \pm 0.3$ M$_{\odot}$ and for the donor star M$_{*} = 1.45 \pm 0.35$ M$_{\odot}$.

We point out that the secondary star in quiescence has an apparent
magnitude m$_V = 17.12$, it has been classified as an F3 IV-F6 IV
sub-giant \citep{Orosz}, and along the line of sight there is an interstellar
absorption A$_V = 3.1 \times \mathrm{E(B-V)} = 4.03$ mag
\citep{Horne}. A sub-giant star of this spectral type has a mean
intrinsic magnitude M$_V \sim 3.2 \pm 0.2$ mag
\citep{Popper,Schmidt-Kaler}, and it would be at a distance
$\mathrm{D} = 950 \pm 150$ pc. Alternatively, if the secondary were a
main sequence star of spectral type F5 V star \citep{Regos}, for an
absorption A$_V$ in the range of 3.3-4.3 mag the distance would be in
the range of 800-1250 pc. However, it may be incorrect to attribute the
absolute magnitudes of isolated stars to secondary stars in X-ray
binaries with the same spectral type.  

Following kind
communications by Beer \& Podsiadlowski (private communication), the
lower luminosity and temperature implied by a distance of $\sim$1 kpc
would require a higher mass ratio with much reduced masses for the
compact object ($\sim$3.2 M$_{\odot}$) and secondary star ($\sim$0.1
M$_{\odot}$). But the absorption spectrum of the secondary seems to rule out 
an M type star of such low mass or a K-type star with mass 
0.6 $\leq$M$_{\odot}$. The spectra of stars with T$_{eff}$$\leq$5000 K have 
different signatures, such as strong molecular bands. This was the case 
in GRS 1915+105 where K band spectroscopy \citep{Greiner3} revealed CO 
molecular bands rendering invalid the classification of the donor as 
a main sequence star.

We point out that there has been analogous uncertainties about the
nature of the secondary in the X-ray binary LMX-3; the proposition
that it is a main sequence star \citep{Cowley,vanderKlis} has recently
been challenged by \cite{Soria2} who argue that it is a
sub-giant. Furthermore, it is not known what could be in GRO J1655-40
the contribution to the optical flux from: 1) the accretion disk
detected in the X-rays at times when it was believed that the source
was in optical quiescence \citep{Garcia}, and 2) possible non-thermal
processes (e.g. synchrotron jets) that may be associated with the
polarization observed in the optical flux \citep{Gliozzi}. In this
context, for the scope of the present study we leave as an open
question the issue on the distance of GRO J1655-40.
 
\section{Space velocity and galactocentric orbit}

As shown below, the parameters of the runaway motion and
galactocentric orbit of GRO J1655-40 are essentially the same for
distances in the range of 0.9-3.2 kpc. The tangential velocity is
 
\begin{equation}
\mathrm{V_t = (25 \pm 3)\,  D_{kpc}\,  km \, s^{-1}}, 
\end{equation}

\noindent where D$_{kpc}$ is the distance in kpc. The radial velocity
with respect to the Sun is -142.4$\pm$1.5 km s$^{-1}$ \citep{Orosz,
Shahbaz}.

Using values of the position, distance, proper motion and radial
velocity, the Galactic orbit of GRO J1655-40 can be computed using a
Galactic gravitational potential model \citep{stdpot}. The velocity
components U, V, and W directed to the Galactic centre, rotation
direction, and north Galactic pole are derived using \cite{Johnson}'s 
equations of transformation, and assuming the sun moves
(U$_{\odot}$,V$_{\odot}$,W$_{\odot}$) = (9, 12, 7) km s$^{-1}$
relative to the local standard of rest (l.s.r.) \citep{Mihalas}. Two
different orbits were computed for two extreme values of the distance: 
for D = 0.9 kpc we obtain (U,V,W) =
($-133\pm2, 27\pm2, 1\pm3$) and for D = 3.2 kpc, (U,V,W) =
($-121\pm18, -33\pm8, 3\pm8$). These two sets of values are rather different 
from the mean values that characterize the kinematics of stars that belong 
to the halo, and the thin and thick disk of the Galaxy \citep{Chiba}.
 The runaway velocities V$_{run}$ were
computed for the two possible extreme distances of 0.9 kpc and 3.2 kpc 
(see Table \ref{tab_par}), after
subtracting the Galactic differential rotation given by the model for
the corresponding position of the source in the Galactic disk.

\begin{table}
{\centering \begin{tabular}{|c|c|c|}
\hline 
D [kpc]&
0.9&
3.2\\
\hline
\hline 
V$_{run}$ [km s$^{-1}$]&
130&
93\\
\hline 
$e$&
0.39&
0.29\\
\hline 
$z_{max}$ [kpc]&
0.05&
0.15\\
\hline 
$r_{max}$ [kpc]&
13.8&
7.2\\
\hline 
$r_{min}$ [kpc]&
6.0&
3.9\\
\hline 
$p$ [M$_{\odot}$ km s$^{-1}$]&
430&
637\\
\hline 
T$_{kin}$ [erg]&
5.6$\times 10^{47}$&
5.9$\times 10^{47}$\\
\hline
\end{tabular}\par}
\caption{\textbf{Parameters of the runaway motion and galactocentric
orbit of GRO J1655-40, for two extreme values of the distance.} 
V$_{run}$ is the runaway velocity in three dimensions after correction
for differential Galactic rotation. $e$ is the eccentricity of the
galactocentric orbit, $Z_{max}$ is the maximal height above the
Galactic plane, and $r_{max}$ and r$_{min}$ are the maximal and
minimal galactocentric distances. The runaway linear momentum $p$ and
kinetic energy T$_{kin}$ of the binary system are computed according
to the masses given in the text.
\label{tab_par}}
\end{table}

The runaway linear momentum $p$ and kinetic energy T$_{kin}$ were
computed assuming M$_{BH}$ = 5.4 $\pm$ 0.3 M$_{\odot}$ and M$_*$ =
1.45 $\pm$ 0.3 M$_{\odot}$ for a distance D = 3.2 kpc \citep{Beer},
and M$_{BH}$ = 3.2 M$_{\odot}$ and M$_*$ = 0.1 M$_{\odot}$ for a
distance D = 0.9 kpc (Beer \& Podsiadlowski, private communication).

\begin{figure*}[htb]
{\centering
\resizebox*{0.48\textwidth}{!}{\rotatebox{0}{\includegraphics{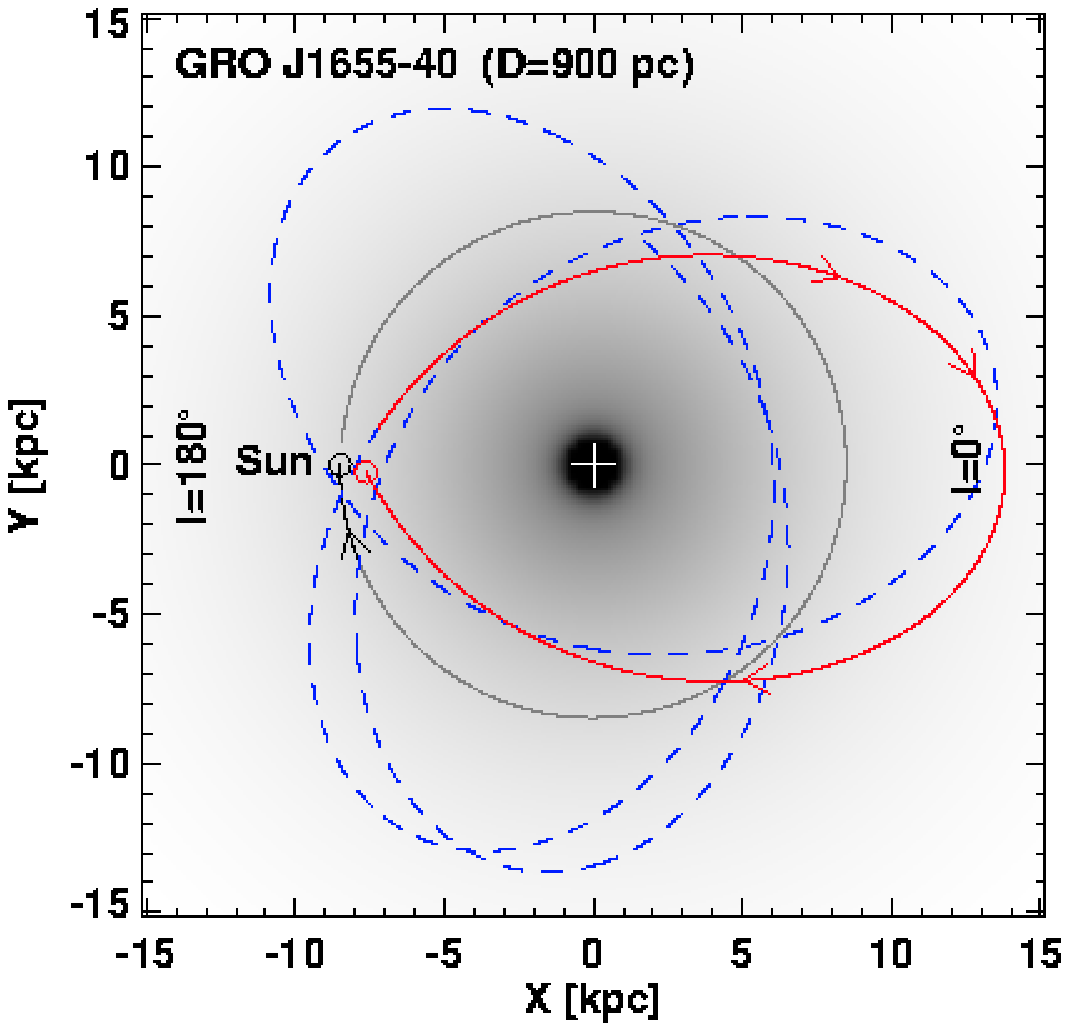}}}
\resizebox*{0.48\textwidth}{!}{\rotatebox{0}{\includegraphics{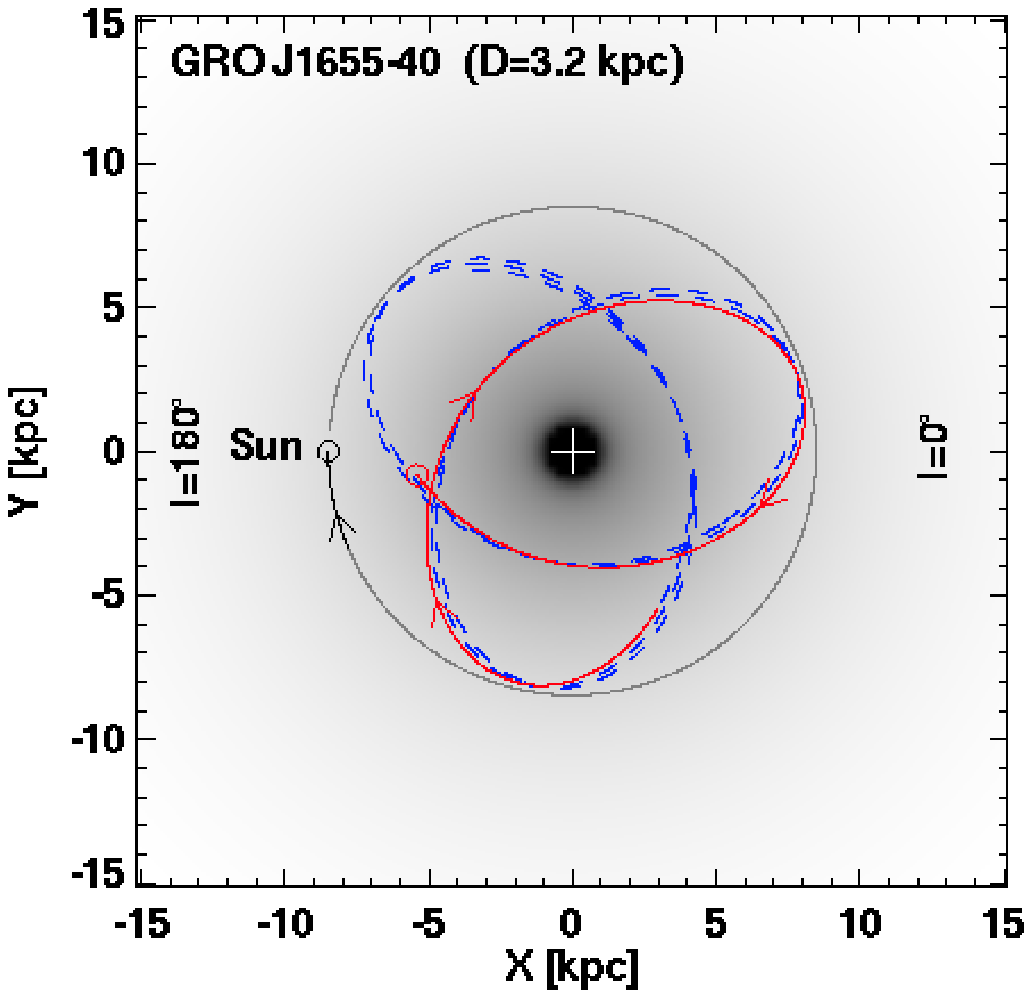}}}
\par}
\caption{\textbf{Galactocentric orbits of GRO J1655-40 viewed from above the
Galactic plane.} The orbital plane is almost parallel to 
the Galactic disk and the source never reaches heights greater than 150 pc. 
On the left is shown the orbit obtained for an heliocentric distance D = 0.9
kpc, and on the right, for D = 3.2 kpc. GRO J1655-40 never penetrates 
the Galactic bulge. Plotted in dashed blue line is
1 Gyr of backwards integration, and in red the past 230 Myrs.
\label{GROorbit}}
\end{figure*}

The parameters of the runaway motion and galactocentric orbit of GRO
J1655-40 are given in Table \ref{tab_par}. In Figure \ref{GROorbit}
are represented the Galactocentric orbits. For a given Galactic potential 
the orbital parameters do not change as a function of time. 
The selection of different Galactic potentials from the current models 
introduce a scatter in the values of the orbital parameters smaller 
than 10\%. In fact, the errors in the parameters listed in Table 1 are 
largely dominated by the uncertainty in the distance.  
For the range of distances
0.9--3.2 kpc, the galactocentric orbit is highly eccentric ($e =
$ 0.29--0.39), the source always moves within a maximum height of 150 pc from
the Galactic plane, and since
the minimum perigalactic distance is 3.9 kpc it never reaches 
the Galactic bulge. 

Clearly, GRO J1655-40 does not move in a halo orbit as XTE J1118+480 \citep{Mirabelnat}. 
GRO J1655-40 must have been born in the Galactic
plane at a galactocentric distance $\geq$ 3.8 kpc.  The runaway linear
momentum and kinetic energy of the binary system are comparable to
those of solitary neutron stars and millisecond pulsars
\citep{Toscano}.

\section{Conclusion}

From the proper motion and radial velocity of GRO J1655-40 -irrespective 
of the uncertainties in the distance- we conclude the following:

\begin{enumerate}

\item{The X-ray binary has a runaway velocity of 112$\pm$18 km s$^{-1}$,  
probably imparted by a natal explosion during the formation of the black hole.  }

\item{The linear momentum of the binary system is 
538 $\pm$ 100 M$_{\odot}$ km s$^{-1}$, which is comparable to that found 
in solitary neutron stars and millisecond pulsars. This suggests that the 
relatively low-mass black hole ($\leq$ 6 M$_{\odot}$) in GRO J1655-40 
may have been formed through a neutron star phase.}

\item{The kinetic energy of the system is (5.8$\pm$0.4) $\times$ 10$^{47}$ 
ergs, namely, $\sim$6 $\times$ 10$^{-4}$ that of a typical supernova.}

\item{The galactocentric orbital plane is almost parallel to the Galactic 
disk and the Galactic pericentre $\geq$ 3 kpc, which indicates that the 
black hole in GRO J1655-40 is the remnant of a massive star born 
in the Galactic disk.} 

\item{Contrary to the binary XTE J1118+480 which harbors a black hole 
born in the halo \citep{Mirabelnat}, 
in the case of GRO J1655-40 there is unambigous kinematical evidence 
that the black hole was born in the disk and received a natal kick, most 
likely from a supernova explosion \citep{Israelian}.

}

\end{enumerate}

\vskip .1in

\begin{acknowledgements}  We thank J. Casares, P. Podsiadlowski,
M.E. Beer, J.-P. Lasota, S. Corbel, P. Benaglia, A. Piatti, and E. Gourgoulhon
for help and discussions on different aspects of this work. J.A.C. and
I.F.M. are members of the Consejo Nacional de Investigaciones
Cientificas y T\'ecnicas of Argentina, and I.R. is a fellow of the
Conselho Nacional de Desenvolvimento Cien\'\i fico e Tecnol\'ogico of
Brazil. I.F.M. acknowledges support from PIP 0049/98 and Fundaci\'on
Antorchas.
\end{acknowledgements}



\end{document}